\documentclass[aps,amssymb,12pt]{revtex4}
\usepackage{graphicx,epsf}
 \advance\oddsidemargin by -1in
 \advance\evensidemargin
 by -1in
 \marginparsep
 0.4cm
 \advance\topmargin by
 -0.0in
 \makeindex

 \newcommand\beq{\begin{equation}}
 \newcommand\eeq{\end{equation}}
 \newcommand\beqn{\begin{eqnarray}}
 \newcommand\eeqn{\end{eqnarray}}
\newcommand{\la}{\langle}
 \newcommand{\ra}{\rangle}
 \newcommand{\ga}{\gamma^*}

\def\fm{\,\mbox{fm}}

\begin{document}

  \title
  {FINAL STATE INTERACTION IN SEMI-INCLUSIVE DIS OFF NUCLEI}
  \author{\large C. Ciofi degli Atti$^a$ and B.Z. Kopeliovich$^{b,c}$}
  \vspace{10mm}
  \affiliation{$^a$ Department of Physics, University of Perugia, and
\\ INFN, Sezione di Perugia, via A. Pascoli, Perugia, I-06100, Italy\\
$^b$Max-Planck-Inst. f\"ur Kernphysik, Postfach 103980, 69029
Heidelberg, Germany\\
$^c$Institut f\"ur Theoretische Physik der Universit\"at, 93040
Regensburg, Germany
}
    \date{\today}

   \begin{abstract}

 The final state interaction (FSI)  in Deep Inelastic Scattering (DIS) of
leptons off a nucleus $A$, due to the propagation of the struck nucleon
debris and its hadronization in the nuclear environment is considered.
The effective cross section of such a partonic system with the nucleons
of the medium and its time dependence are estimated, for different values
of the Bjorken scaling variable, on the basis of a model which takes into
account both the production of hadrons due to the breaking of the color
string, which is formed after a quark is knocked out off a bound nucleon,
as well as the production of hadrons originating from gluon radiation. It
is shown that the interaction, the evolution and the hadronization of the
partonic system in the nuclear environment can be thoroughly investigated
by a new type of semi-inclusive process, denoted $A(e,e'(A-1))X$, in
which the scattered lepton is detected in coincidence with a heavy
nuclear fragment, namely a nucleus $A-1$ in low energy and momentum
states.  As a matter of fact, if FSI is disregarded, the momentum
distribution of $A-1$ is directly related to the momentum distribution of
the nucleon before $\gamma^*$ absorption, i.e. the same quantity which
appears in the conventional $A(e,e'N)X$ process, where $N$ denotes a
nucleon. The rescattering of the struck nucleon debris with the medium
damps and distorts the momentum distributions of $A-1$ in a way which is
very sensitive to the details of effective cross section of the debris
with the nucleons of the medium. The total cross section of the process
$A(e,e'(A-1))X$ on $^4He$, $^{16}O$, and $^{40}Ca$, related to the
probability that after a target nucleon experiences a DIS process, the
recoiling $A-1$ nucleus remains intact in spite of the strong FSI, is
evaluated, and the distorted momentum distribution of the recoiling $A-1$
system is obtained. It is shown that both quantities are very sensitive
to the details of the early stage of hadronization of the nucleon debris
in the nuclear medium.

   \end{abstract}
 \maketitle


  \vspace{0.5cm}

  \newpage
  \section{Introduction}

Lepton scattering off nuclei in the Deep Inelastic Scattering (DIS)
regime represents a powerful tool to investigate a wide range of physics
processes related to strong interaction physics which are more difficult
to study in lepton scattering off free nucleons. As a matter of fact, in
lepton nucleus scattering  the interaction of the debris of the struck 
nucleon with
the nuclear medium during the hadronization processes, could provide in
principle valuable information on the space time structure of the hadron
formation mechanism (see e.g. \cite{hera}). Among others, we should
stress two main motivations for thoroughly investigating quark
interaction in the hadronic medium:
 \begin{enumerate}
 \item the first one, as already pointed out, is related to the
possibility to understand the very mechanism of the formation time and
hadronization;
  \item the second one stems from the obvious necessity, once a workable
model for quark propagation and hadronization is developed, to apply to
the treatment the Final State Interaction (FSI) in various
processes involving nuclei, like e.g., ultra relativistic heavy ion
collisions, aimed at observing a possible quark-gluon plasma formation,
and semi inclusive lepton DIS scattering processes, aimed at
investigating possible distortions of the nucleon structure function of a
bound nucleon (EMC effect).
 \end{enumerate} 
 It is the aim of this paper to update a comprehensive and workable model
to treat the propagation and re-interaction in the medium of a nucleon 
debris
produced in a DIS process off a bound nucleon \cite{knp}, and to apply it
to a recently proposed semi-inclusive DIS process on nuclei \cite{cks},
which could be useful not only to investigate in a more detailed way the
mechanism of formation length and hadronization, but also to obtain more
reliable information on EMC-type effects.

To date, the information on hadron formation length comes mainly from the
measurement of the multiplicity ratio of the lepto-produced hadrons in
semi inclusive $A(e,e'h)X$ processes (\cite{osborne,ashman,HERMES}),
whose interpretation on the basis of the quark re-interaction model we
consider in this paper \cite{knp} appeared to be very convincing (see
e.g. \cite{muccifora} for a discussion of recent HERMES data). However,
it should also be pointed out, that more exclusive processes, e.g.  of
the type we are going to discuss, though difficult to perform, could in
principle provide more direct information on quark re-interaction and
hadronization mechanisms. As for the EMC-type effects, in spite of many
experimental and theoretical efforts (for a recent review see
\cite{arneodo}), the origin of the nuclear EMC effect has not yet been
fully clarified, and the problem as to whether, and to which extent, the
quark distributions of nucleons undergo deformations due to the nuclear
medium, remains open. This is why various semi-inclusive experiments in
which another particle is detected in coincidence with the scattered
electron have been proposed. Most of theoretical studies in this field
concentrated on the process $D(e,e'N)X$, where $D$ denotes the deuteron,
$N$ a nucleon, and $X$ the undetected hadronic state. Current theoretical
models of this process are based upon the {\it impulse approximation}
($IA$) (or the {\it the spectator model} ), according to which: i) $X$
results from DIS off  one of the two nucleons in the deuteron, ii) the second nucleon $N$
recoils without interacting with $X$ and is detected in coincidence with
the scattered electron (for an exhaustive review see \cite{fs}). The
model has been improved by considering that the detected nucleon could
also originate from quark hadronization \cite{dieper,ciofisim}, and has
also been extended to complex nuclei by considering the process
$A(e,e'N)X$, by assuming that DIS occurs on a nucleon of a correlated
pair, with the second nucleon $N$ recoiling and detected in coincidence
with the scattered electron \cite{ciofisim}.

 In all above calculations, however, the nucleon debris created by the
virtual photon is assumed to propagate without re-interacting with the
spectator nucleus, which, therefore, always remains intact, an assumption
which, at first sight, might appear unjustified.

 As a matter of fact, DIS scattering off a bound nucleon results in the
production of a multi-particle final state with an effective mass squared
equal to ,
 \begin{eqnarray}
s'&&\simeq m_N^2-Q^2 + 2\,m_N\,\nu -
2\,\sqrt{\nu^2-Q^2}\,p_L\nonumber\\
&&=Q^2(\frac{1}{x}-1)+m_N^2 -2|{\bf q}| p_L
\label{b.1}
 \end{eqnarray}
 where $Q^2={\bf q}^2 -{\nu}^2$ is the four- momentum transfer, $\nu$ the
virtual photon energy in the rest frame of the nucleus, $p_L$ the
longitudinal Fermi momentum of the nucleon relative to the direction of
the virtual photon (${\bf q}$ $\parallel z$), $m_N$ the nucleon mass, and
$x=\displaystyle\frac {Q^2}{2m_N\nu}$ the Bjorken scaling variable (we
neglect here the binding energy of the nucleon). At high energies and far
from the quasi-elastic region ($x\approx 1$), the effective mass is large,
$\sqrt{s'} \gg m_N$, and one could expect production of many particles
which can interact traveling through the nucleus. This would
substantially suppress the probability for the spectator nucleus to remain
intact. However, the process of multi-particle production has a specific
space-time development, and it turns out that not so many particles have
a chance to be created inside the nucleus.

Recently \cite{cks}, a new type of semi-inclusive process on complex
nuclei has been considered, namely the process $A(e,e'(A-1))X$, in which
DIS occurs on a mean-field, low-momentum nucleon, and the nucleus
$(A-1)$, recoiling with low momentum and low excitation energy, is
detected in coincidence with the scattered electron. Within the IA, it
has been shown that such a process exhibits a series of very interesting
features which could in principle provide useful insight on the nature
and the relevance of quark FSI in DIS off nuclei, the validity of the
spectator mechanism, and the medium induced modifications of the nucleon
structure function. In the present paper we go beyond the IA by
considering the effects of the quark re-interaction in order to clarify:
i) to which extent the conclusions reached in Ref. \cite{cks} will be
affected by the FSI, and ii) if, and to which extent, the process is
sensitive to the details of quark hadronization in nuclear environment.

 Our paper is organized as follows: in Section 2 the formalism of FSI is
presented; the application of the theory to the process $A(e,e'(A-1))X$
is illustrated in Section 3; the Summary and Conclusion is given in
Section 4.

\section{Final State Interaction and Hadronization in semi-inclusive
processes} 

\subsection{Coherence time for particle production. The color string
model}

After a quark is knocked out off a bound nucleon by the virtual photon, a
color field is stretched between the quark and the remnants of the
nucleon. In the color flux tube model \cite{cnn,g},  one assumes that this
process is adiabatic, {\it i.e.} gluon radiation is neglected  while the
color field is squeezed by the QCD vacuum to a color tube of a constant
cross area.  Assuming that the transverse dimension of the tube is much
less than its length one can call it color string.

The important parameter of the model is the string tension, {\it i.e.}
the energy density per unit of length, which is related
to the hadronic mass spectrum \cite{cnn},
\beq
\kappa = \frac{1}{2 \pi \alpha_R^\prime}\simeq \frac{1 GeV}{fm}\ ,
\label{b.1a}
\eeq
where $\alpha^\prime_R\approx 0.9\,GeV^{-2}$ is the slope of the
leading Regge trajectories.

The energy of the leading quark degrades with a constant rate,
$dE/dz=-\kappa$, ($z$ is the longitudinal coordinate) which is invariant
relative Lorentz boosts. At the same time, the rest of the nucleon ({\it
e.g.} a diquark) speeds up with the same acceleration, so the total
momentum and energy of the debris of the nucleon remain constant. The
string itself carries only energy, but no longitudinal momentum.

Naively one might think about a long string stretched across the nucleus.
However, the slow end of the string is accelerated and reaches soon the
speed of the light. It turns out that the maximal length of the string
in the rest frame of the target nucleon is,
 \beq
L_{max}=\frac{m_{qq}}{\kappa}\ ,
\label{b.2}
 \eeq
 where $m_{qq}$ is the mass of the rest of the nucleon, which we
conventionally call a diquark. Assuming this mass to be less than the
mass of the nucleon, one gets $L_{max} < 1\,fm$. Thus the string
propagating through the nucleus is rather a short object in the nuclear
rest frame.  Moreover, subsequent decays of the string via spontaneous
$\bar qq$ pair production from vacuum make the string even shorter(see
below).

Since the interaction cross section of a high-energy colorless object in
QCD depends only on its transverse size, one can assume that for such a
string it should be of the order of nucleon-nucleon total cross section.

An important phenomenon related to the string evolution is the
spontaneous creation of quark-antiquark pairs from vacuum via the
Schwinger mechanism. Since the string potential is a linear function of
the distance, a created $\bar qq$ pair completely screens the external
fields in between, therefore, it breaks the string into two pieces. If
so, the interaction cross section may nearly double compared to one
string. This is important and must be taken into account.

The probability $W(t)$ for a string to create no quark
pairs since its origin is given by
\beq
W(t) = {\rm exp}\left[
-\,w\,\int\limits_0^t dt'\,L(t')\right]\ ,
\label{b.3}
\eeq
where $w$ is the  probability rate
to create a light $\bar qq$ pair within a unit of length of
the string and  a unit of time ($t=z$, in natural units), and  $L(t)$ is the
time dependent
length of the string. Note that Eq.~(\ref{b.3}) is
invariant relative to longitudinal Lorentz boosts.

The key parameter $w$ can be estimated either using the
Schwinger formula, or by calculating the decay width of
heavy resonances \cite{cnn,g},
both ways giving $w\approx 2\,fm^{-2}$.
One can also evaluate $w$ from the momentum distribution
of the  recoil protons in the reaction $pp\to pX$ \cite{kn1,kn2}.
If the final proton stays in the fragmentation region
of the target,  it acquires momentum due to the  acceleration
by the string up to the moment $\Delta t$
of the first string breaking.

This moment is determined according to (\ref{b.3}) by the condition
 \beq
{1\over2}\,w\,\Delta t^2 \approx 1
\label{b.4}
 \eeq
 The momentum which the proton gets during this time interval,
$p=\kappa\,\Delta t$, is related to the Feynman $x_F$ of the
leading proton (in the anti-laboratory frame),
$p=m_N\,(1-x_F^2)/2x_F$. The mean value of
$x_F$ is known from data, $\la x_F\ra\approx 0.5$.
Therefore, $\la p\ra \approx 1\,GeV/c$, and
$w = 2/\Delta t^2\approx 2\,\kappa^2/p^2\approx 2\,fm^{-2}$.
Thus, all of these estimates converge at about the same  value.

From this consideration we found that the mean time of breaking of the
string after its production is $\Delta t\approx 1\,fm$. Since after each
breaking the leading piece of the string becomes twice shorter (in the
lab frame), the time interval up to the next breaking, and
correspondingly the momentum of the next produced particle, double. This
is the way how string decay produces sequences of hadrons whose momenta
are ordered in geometrical progression, corresponding to a plateau in
rapidity scale. This bunch of particles with multiplicity rising with
time propagates through nuclear matter with increasing probability to
interact, i.e. to break up the recoiling nucleus.  The effective
interaction cross section of the partonic system developing in nuclear
matter rises as function of time as,
 \beq
\sigma_{eff}(t)=\sigma^{NN}_{tot} +
\sigma^{M N}_{tot}\,n_{M}(t)\ ,
\label{b.5}
 \eeq
 where
 \beq
n_{M}(t)=
\frac{{\rm ln}(1+t/\Delta t)}
{{\rm ln}2}\ ,
\label{b.5a}
 \eeq
 and we have assumed that the string at the early stage before the first
breaking interacts with the nucleon cross section. Subsequent decays of
the string are assumed to be, $str \to B + str \to B + M + str \to B +
2\,M + str {\ldots} $, where $B$ and $M$ are baryon and meson
respectively. All other produced particles are mesons which are assumed
to interact with the pionic cross section, $\sigma^{MN}_{tot}=\sigma^{\pi
N}_{tot}$. We also assume that they decay predominantly outside the
nucleus. The latter assumptions is well justified for few-GeV mesons.

The effective cross section (\ref{b.5}) grows logarithmically with time
and in a long time interval, $t\,\propto\,E_q$, when the hadronization is
completed, reaches the maximal value,
 \beq
\sigma_{max} \approx \sigma^{NN}_{tot} +
\sigma^{\pi N}_{tot}\,\la n_{M}\ra\ ,
\label{b.5b}
 \eeq
 where $\la n_{M}\ra$ is the observed mean multiplicity of produced
mesons. If the energy of the quark initiating the jet is sufficiently
high, the late stage of string hadronization happens outside the nucleus.

\subsection{Gluon bremsstrahlung}

An intuitive QED analogy for the string model would be a capacitor whose
plates are much larger than the distance between them. These plates
moving apart are losing energy mainly for creation of the static electric
field, while photon radiation from the edges is a small correction. This
analogy gives a hint to the restrictions for the application of the
string model. Probably, soft inelastic interactions can be treated this
way, although the size of a constituent quark and the length of the
color-flux tube $\sim 1 fm$ (up to the first break) are of the same
order, and gluon radiation might be an important correction. However, a
quark knocked out by a highly virtual photon in DIS has size $\sim
1/Q$, much smaller than the length of the string.  This is apparently a
situation when the "edge effect" of gluon radiation plays major role
\cite{knp}.

 In order to estimate the effect of gluon radiation one can rely upon
perturbative QCD methods \cite{knp,kst1,kst2}. We are interested in the
time-dependence of the amount of radiated gluons, therefore the coherence
time of radiation is important. It depends on the quark energy which is
about the energy of the incident virtual photon $E_q\approx \nu$ , the
transverse momentum $k_T$ of the gluon, and the fraction $\alpha$ of the
quark light-cone momentum it carries \cite{knp,kst1},
 \beq
t_c=\frac{2\,E_q\,\alpha\,(1-\alpha)}
{k_T^2 + \alpha^2\,m_q^2}\ ,
\label{bb.1}
\eeq
 where $m_q$ is the quark mass which is not important for further 
estimates.

The mean number of gluons which lose coherence and are radiated
during the time interval $t$ is given by \cite{knp,hk},
  \beq
n_G(t)=
\int\limits_{\lambda^2}^{Q^2} dk_T^2
\int\limits_{k_T/E_q}^1 d\alpha\,\frac{dn_G}{dk^2_T\,d\alpha}\,
  \Theta(t-t_r)\ ,
  \label{bb.2}
  \eeq
 where the number of radiated gluons as a function
of $\alpha$ and $\vec k_T$ reads \cite{n,knp},
 \beq
\frac{dn_G}{d\alpha\,dk_T^2} =
\frac{4\alpha_s(k_T^2)}{3\,\pi}\;\frac{1}{\alpha\,k_T^2}
\label{bb.3}
 \eeq
 Here $\alpha_s(k_T^2)=4\,\pi/9\,ln(k_T^2/\Lambda_{QCD}^2)$ is the
leading order QCD running coupling. We use the approximation of soft
radiation, $\alpha\ll 1$ and $k_T^2\ll Q^2$.  Since $dn_G/dk_T^2\propto
1/k^4$ at $k_T^2\gg Q^2$ \cite{bg} we can use $Q^2$ as the ultra-violet
cut off in the $k_T^2$ integration in eq. (\ref{bb.2}).

 To avoid double counting we assume that at $Q < \lambda$ the string
fragmentation mechanism dominates the nonperturbative dynamics of
particle production. For this reason we introduced in (\ref{bb.2}) an
infrared cut-off $\lambda^2$ for the $k_T^2$-integration, which should be
taken of the order of the semi-hard scale characterizing gluon radiation.
This scale was fixed in \cite{kst2} at $\lambda=0.65\,GeV$ from data on
diffractive gluon radiation (the triple-Pomeron mechanism) in soft
hadronic collisions. The relatively large value of $\lambda$
corresponding to a semi-hard scale, is dictated by the experimentally
observed smallness of the cross section of single diffraction, $pp
\rightarrow pX$ to states of large effective mass $M_X$. The parameter
$\lambda$ controls the size of the virtual gluon clouds surrounding he
valence quarks. Although the mean radius of the cloud $r_0 \simeq 0.3 fm$
might seem to be small, it goes well along with the radius of gluon-gluon
correlation, calculated on the lattice \cite{dg}, or with the instanton
phenomenology \cite{shuryak} which gives similar estimates.  The
appearance of such a semi-hard scale has presumably a nonperturbative
origin and was interpreted in \cite{kst2} as a result of a strong
nonperturbative light-cone potential between the parent quark and gluon.

After integration in (\ref{bb.2}) we find that multiplicity of radiated
gluons rises with time differently dependent on whether $t$ is smaller or
larger than $t_0=1/(m_M\,x_{Bj})= 0.2\fm/x_{Bj}$. At $t < t_0$,
 \beq
n_G(t) = \frac{16}{27}\,\left\{
{\rm ln}\left(\frac{Q}{\lambda}\right)\,+\,
{\rm ln}\left(\frac{t\,\Lambda_{QCD}}{2}
\right)\,{\rm ln}\left[\frac{{\rm ln}(Q/\Lambda_{QCD})}
{{\rm ln}(\lambda/\Lambda_{QCD}}\right]\right\}\ .
\label{bb.4a}
 \eeq
 At $t > t_0$ the $t$-dependence starts leveling off,
 \beqn
n_G(t) &=& \frac{16}{27}\,\left\{
{\rm ln}\left(\frac{Q}{\lambda}
\,\frac{t}{t_0}\right)\,+\,
{\rm ln}\left(\frac{t\,\Lambda_{QCD}}{2}
\right)\,{\rm ln}\left[\frac{{\rm ln}(Q/\Lambda_{QCD}
\sqrt{t_0/t})}
{{\rm ln}(\lambda/\Lambda_{QCD}}\right]
\right.\nonumber\\ &+& \left.
{\rm ln}\left(\frac{Q^2\,t_0}{2\,\Lambda_{QCD}}
\right)\,{\rm ln}\left[\frac{{\rm ln}(Q/\Lambda_{QCD}}
 {{\rm ln}(Q/\Lambda_{QCD}\,\sqrt{t_0/t})}\right]\right\}
\ ,
\label{bb.4b}
 \eeqn
 and reaches the maximal constant value when the time exceeds the full
hadronization time, $t > t_0\,Q^2/\lambda^2$,
 \beq
n_G^{max} = \frac{16}{27}\,\left\{
-\,{\rm ln}\left(\frac{Q}{\lambda}\right)\,+\,
{\rm ln}\left(\frac{Q^2\,t_0}{2\,\Lambda_{QCD}}
\right)\,{\rm ln}\left[\frac{{\rm ln}(Q/\Lambda_{QCD})}
{{\rm ln}(\lambda/\Lambda_{QCD}}\right]\right\}\ .
\label{bb.4c}
 \eeq
 Apparently, all the three regimes Eqs.~(\ref{bb.4a})--(\ref{bb.4c})  
match.

Thus, we have arrived at a similar logarithmic growth of the amount of
the produced particles with time as in the string model.  Note that no
gluons is radiated at $Q \rightarrow \lambda$, which is reasonable since
at this value the onset of the nonperturbative dynamics of gluon
radiation is expected \cite{kst2}. At $Q > \lambda$ the amount of gluons
slowly rises with $Q$. This observation does not contradict the fact that
gluons provide the main contribution to the energy loss at large 
values of $Q^2$ \cite{knp}. Although the amount of hard (high $k_T$)
gluons is small, each of them takes away a large value of energy ,
$\omega > k_T^2t$.

\begin{figure}[tbh] 
\includegraphics{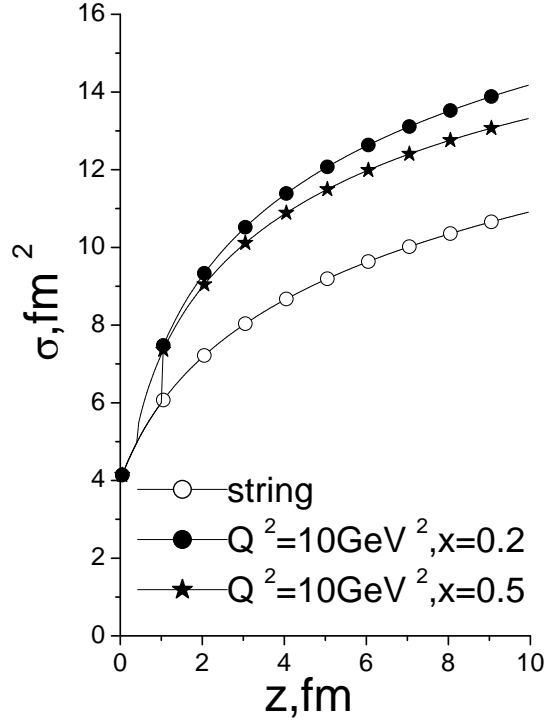}
\begin{center}                  
\vspace{10cm}
\parbox{13cm}
{\caption[Delta]
{The effective quark debris-nucleon cross section {\it vs} the
longitudinal co-ordinate $z$ in correspondence of $Q^2=10 GeV^2$ and two
values of the Bjorken scaling variable $x$. The open dots
corresponds to the string model (Eq.\ref{b.5}), and the full dots and
stars represents the total $\sigma_{eff}$ i.e. the sum of the string and
gluon radiation contributions (Eq. \ref{bb.6}).
 }
\label{Fig3}}
\end{center}
\end{figure}

 One can replace each radiated gluon by a color octet $\bar qq$ pair with
accuracy $1/N_c^2$, and rearrange these pairs to form colorless $\bar qq$
dipoles. If to treat those dipoles as produced mesons, the effective
cross section (\ref{b.5}) get an extra contribution,
 \beq
\sigma_{eff}(t)=\sigma^{NN}_{tot} +
\sigma^{\pi N}_{tot}\Bigl[n_M(t) +
n_G(t)\Bigr]\ ,
\label{bb.6}
 \eeq
 which smoothly switches into the expression (\ref{b.5a}) at
$Q\rightarrow \lambda$.  The effective absorption cross section
$\sigma_{eff}$ as function of time is depicted in Fig.~\ref{Fig3} for
different values of $x_{Bj}$.

\section{The semi-inclusive process $\mathbf{A(e,e'(A-1))X}$}

\subsection{The Impulse Approximation}

The process we are going to consider is the one in which $\gamma^*$,
interacting at high $Q^2$ with a quark of a mean field nucleon (to be
labeled by $``1"$, and considered to be a proton), having four-momentum $p_1 \equiv(p_{10},\vec {p_1})$,
with $\vec {p_1} \equiv - \vec P_{A-1} $, creates a nucleon  
debris which propagates in the $A-1$ nucleon system, which recoils with low
momentum ${\vec P}_{A-1}$ and low excitation energy, and is detected in
coincidence with the scattered electron. We are interested in the
propagation and interaction of the quark debris with the nuclear
environment, but, to better understand the problem, we first analyze the
{\it Impulse Approximation} ($IA$), when any kind of FSI is disregarded.
In one-photon-exchange approximation, the differential cross section in
the laboratory system has the following form \cite{cks}
   \begin{eqnarray}
   &&\!\!\!\!\!\! \sigma^A_1 (x,Q^2,\vec P_{A-1})\equiv\sigma^A_1=
  \frac{d\sigma^A}{d x d Q^2  d \vec   P_{A-1}}\nonumber\\&&
   =
   K^A( x,Q^2,y_A,z_1^{(A)}) z_1^{(A)}
   F_2^{N/A}(x_A,Q^2,p_1^2)\,P^A(E,|{\vec P}_{A-1}|),
   \label{crosa-1}
   \end{eqnarray}
   where $Q^2 =-q^2= -(k_e-k_e')^2 = \vec q^{\,\,2} - \nu^2=4 E_e E_e^{'}
sin^2 {\theta \over 2}$ is the four-momentum transfer (with $\vec q =
\vec k_e - \vec k_{e}^{'}$, $\nu= { E}_e - {E}_e' $ and $ \theta \equiv
\theta_{\widehat{\vec k_e \vec k_{e}^{'}}}$), $ x = Q^2/2M\nu $ is the
Bjorken scaling variable, $p_1 \equiv(p_{10},\vec {p_1})$, with $\vec
{p_1} \equiv - \vec P_{A-1} $, is the four momentum of the nucleon before
interaction with $\ga$, $F_2^{N/A}$ the DIS structure function of a
nucleon bound in nucleus $A$, and $K^A( x,Q^2,y_A, z_1^{(A)}) $ the
following kinematic factor
   \begin{eqnarray}
   &&
   K^A( x,Q^2,y_A,z_1^{(A)})=
   \frac{4\alpha^2}{Q^4} \frac {\pi}{x}\cdot
     \left( \frac{y}{y_A}\right)^2
   \left[\frac{y_{A}^2}{2} + (1-y_A) -
   \frac{p_1^2x_{Bj}^2 y_A^2}{z_1^{(A)2}Q^2}\right ]~,
   \label{ka}
   \end{eqnarray}
with
\begin{eqnarray}
&&
y=\nu/ {E}_e ~, \,\,\,\,  y_A = (p_1\cdot q)/(p_1\cdot k_e)
\label{ydef}\\
&&
 x_A = {x_{Bj} \over z_1^{(A)}}, \quad
z_1^{(A)} = {p_1 \cdot q \over M \nu}~.
\label{adef}
\end{eqnarray}

 In Eq. \ref{crosa-1}, the quantity $P(E,|{\vec P}_{A-1}|)$ is the
Nucleon Spectral Function
 \beq
P^A(E,|{\vec P}_{A-1}|) = \sum_{f}|\langle{\vec
P}_{A-1},{\Psi_{A-1}^f}|{\Psi_A^0}\rangle|^2
\delta \bigl(E-(E_{min}+E_{A-1}^f)\bigr)
\label{Pke}
 \eeq
 where $\Psi_A^0$ and $\Psi_{A-1}^f$ are the wave functions of the target
nucleus and the final nucleus $(A-1)$ in excited intrinsic state $f$,
respectively, $E=E_{min}+E_{A-1}^f$ is the nucleon removal energy,  i.e. the
energy required to remove a nucleon from the target, leaving the $A-1$
nucleus with excitation energy $E_{A-1}^f$, and, eventually,  $E_{min} =
M+M_{A-1}-M_A$.

From now-on, the overlap appearing in Eq.~(\ref{Pke}) will be called the
{\it transition form factor} of the process and will be denoted as
follows
 \begin{eqnarray}
&&F_{A-1,A}^f({\vec P_{A-1}})\equiv\langle{\vec
P}_{A-1},\Psi_{A-1}^f|\Psi_A^0\rangle \nonumber\\
&&=\int e^{i{\vec P}_{A-1}{\vec r}_1}
{\Psi_{A-1}^f}({\vec r}_2\dots{\vec r}_A)\Psi_A^0({\vec r}_1, {\vec
r}_2\dots{\vec r}_A)
\delta ({\sum_{j=1}^A {\vec r}_j})
\prod_{i=1}^Ad{\vec r}_i,
\label{overlap1}
\end{eqnarray}

Because of energy conservation
\beq
\nu+M_A=P_{(A-1)0}+p_{x0}\,
\label{energycons}
\eeq
 where $p_{x0}$ is the total energy of the debris, the removal energy can
be written in the following way
 \beq
E=\nu+M-p_{x0}
\label{missingen}
 \eeq
 if the total energy of the system $A-1$ is approximated by its
non-relativistic expression and the recoil energy disregarded.

As is well known \cite{cps}, the integral of the spectral function over
the removal energy $E$ defines the (undistorted) momentum distributions
 \beqn
n^A(|{\vec p}|)&=& \int e^{-i{\vec p} ({\vec r} - {\vec r}\,')}
\rho ({\vec r},{\vec r}\,')d{\vec r}d{\vec r\,'}\nonumber\\
&=&\int dE P^A(E,|{\vec p}|) = n_0^A({\vec p}) + n_1^A({\vec p})\ ,
\label{nk}
 \eeqn
 where
 \beq
\rho ({\vec r},{\vec r}\,')=\int {{\Psi_A^0}^{*}({\vec r},{\vec
r}_2\dots{\vec r}_A)\Psi_A^0({\vec r}\,',
{\vec r}_2\dots{\vec r}_A)}\prod_{i=2}^Ad{\vec r}_i\ ,
\label{rho}
 \eeq
 is the one-body mixed density matrix. In Eq. \ref{nk},  $n_0^A$ represents  the mean-field uncorrelated
momentum distribution arising  from the summation over the discrete
hole states of the final system, whereas  $n_1^A$ is  the correlated momentum
distribution resulting from the summation over high excitation states
of the final system, which originate  from nucleon-nucleon correlations.

\begin{figure}[tbh] 
\includegraphics{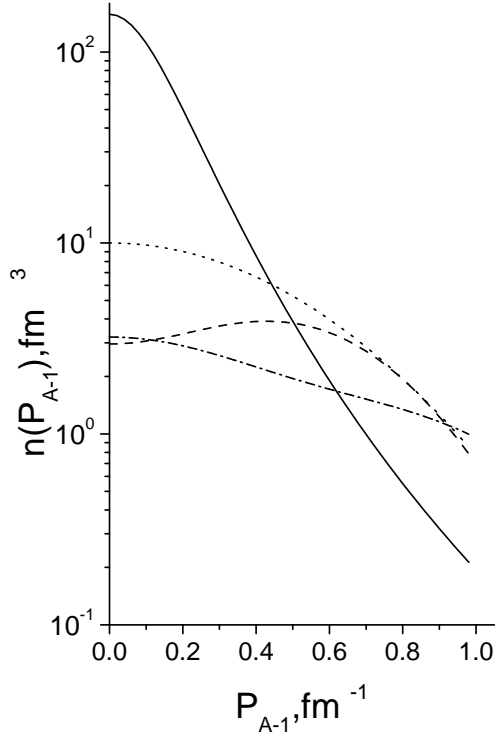}
\begin{center}                  
\vspace{10cm}
\parbox{13cm}
{\caption[Delta]
 {The proton momentum distributions in $^2H$ (full), $^4He$ (dotted),
$^{16}O$ (dashed) and $^{40}Ca$ (dot-dashed) calculated using realistic
nucleon-nucleon interactions(see \cite{cks} for original references).}
 \label{Fig1}}
\end{center}
\end{figure}

In this paper we will consider semi-inclusive processes, when the
cross section (\ref{crosa-1}) is integrated over the removal energy $E$,
at fixed value of ${\vec P}_{A-1}$. Thus, owing to
 \beq
\sum_{f}{\Psi_{A-1}^f}^*({\vec r}_2\,'\dots{\vec
r}_A\,')\Psi_{A-1}^f({\vec r}_2
\dots{\vec r}_A)=\prod_{j=2}^A\delta({\vec r}_j-{\vec r}_j\,')\ ,
\label{closure}
 \eeq
 the cross section (\ref{crosa-1}) becomes directly proportional to the
momentum distribution $n(|{\vec P}_{A-1}|)^A$. However, since, as
previously stated, we will only consider the formation of a low momentum
and low excitation energy $A-1$ system, the summation over $f$ is
effectively limited to the hole states $\alpha$ of the initial nucleus,
which means that the only relevant quantity is the low-momentum part of
 \beq
  n^A_0(|{\vec P}_{A-1}|) =
\displaystyle\sum\limits_{\alpha
   < F}|F_{A,A-1}^{\alpha
   < F}(|\vec P_{A-1}|)|^2\ ,
\label{n0}
 \eeq
 where the sum extends over the states below the Fermi sea occupied in
the ground state. Following \cite{cks}, let us briefly discuss the $A$
dependence of the cross section (\ref{crosa-1}). In this respect it
should be pointed out that nuclear effects are not only generated by the
nucleon momentum distribution $n_0^A(|\vec P_{A-1}|)$, but also by the
quantities $y_A$ and $z_1^{(A)}$, which differ from the corresponding
quantities for a free nucleon ($y=\nu/{E}_e$ and $z_1^{(N)}=1$)  if the
off mass shell of the nucleon ($p_1^2\neq M^2$ ) generated by nuclear
binding is taken into account. However such a dependence is not only very
small, but it completely disappears in the ratio between the cross
sections from nuclei $A$ and $A'$,
 \beqn
 R_{Bj}(x_{Bj}/z_1^{(A)},Q^2,|\vec P_{A-1}|,A,A') & = &
 \frac {F_2^{N/A}(x_{Bj}/z_1^{(A)},Q^2)}
         {F_2^{N/A'}(x_{Bj}/z_1^{(A')},Q^2)}
\frac{n_0^A(|\vec P_{A-1}|)}{n_0^{A'}(|\vec P_{A-1}|)}
\rightarrow
\\
& \rightarrow &
\frac{n_0^A(|\vec P_{A-1}|)}{n_0^{A'}(|\vec P_{A-1}|)}\equiv
R_{A,A'}(| {\vec P_{A-1}}\ |),
  \label{eq5}
 \eeqn
 which means that in the Bjorken limit the $A$ dependence of the ratio
$R$ is entirely governed by the $A$ dependence of the nucleon momentum
distribution $n_0^A(|\vec P_{A-1}|)$ in nuclei $A$ and $A'$. Since, as shown in Fig.~\ref{Fig1}, $n_0^A$
exhibits a strong $A$ dependence for low values of $|\vec P_{A-1}|$, a
plot of $R$ versus $|\vec P_{A-1}|$  should
follow the behavior of $n_0^A(|\vec P_{A-1}|)$ in nuclei $A$ and $A'$
and the experimental observation of such a behavior would represent a
stringent test of the spectator mechanism independently of the model for
$F_2^{N/A}$.

\begin{figure}[tbh] 
\includegraphics{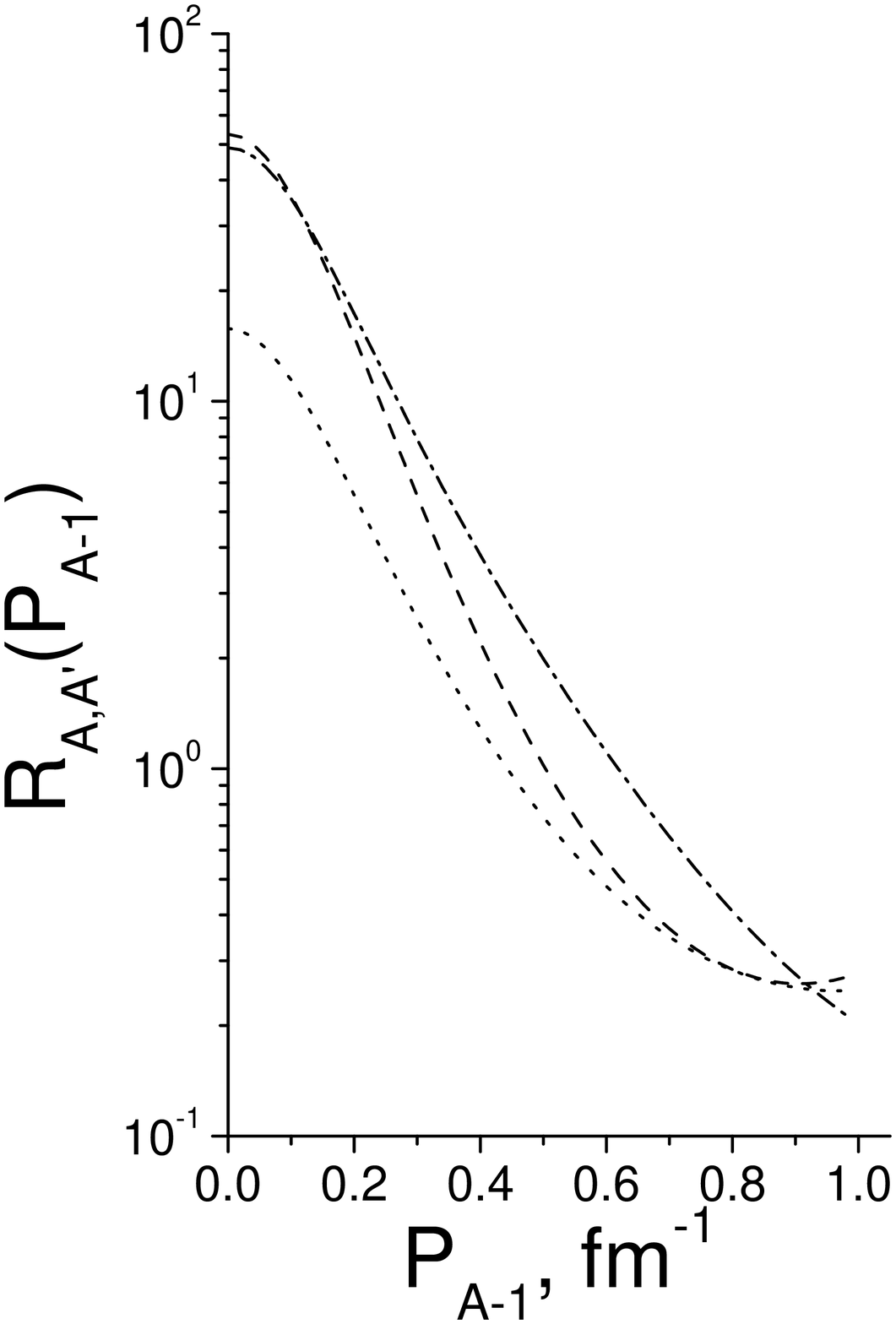}
\begin{center}                  
\vspace{10cm}
\parbox{13cm}
{\caption[Delta]
{ The ratio $R_{A,A'}(P_{A-1})$ (Eq. (\ref{eq5}) with $P_{A-1} \equiv |{\vec P}_{A-1}|$) corresponding to
$A=2$ and $A'=4$ (dotted), $A'=16$ (dashed) and $A'=40$ (dot-dashed).}
\label{Fig2}}
\end{center}
 \end{figure}

 The expected behavior of the ratio
Eq.~(\ref{eq5}) for $A=2$ and different values of $A'$ is presented in Fig.~\ref{Fig2} .
 These results clearly show that the observation of recoil nuclei in the
ground state, with a $|\vec P_{A-1}|$-dependence similar to the one
predicted by the momentum distributions, would represent, on one hand, an
indication that the FSI between the lepto-produced hadronic states and
the nuclear medium is such as to leave intact part of the the final
$(A-1)$ nuclei. One could hope that the number of the detected nuclei
together with the momentum dependence of the cross section could provide
information on the nature and the details of the hadronization mechanism
in a more sensitive way than e. g.  the energy transfer dependence of
forward hadro-production in the process $A(e,e'h)X$
\cite{osborne,HERMES}.

\subsection{Final state interaction and hadronization}

In this Section we are going to consider the modifications of the cross
section induced by the FSI of the nucleon debris produced in the DIS on a
bound nucleon. In this case, the Spectral Function has to be replaced by
the {\it Distorted} ($D$) Spectral Function, which can be written in the
following way
 \beq
P_D^A(E,{\vec P}_{A-1})=\sum_{f}|F_{A-1,A}^{f,D}({\vec
P}_{A-1})|^2\delta(E-(E_{min}+E_{A-1}^{f}))\ ,
\label{Pempm}
\end{equation}
 where the transition form factor is
 \beqn
&&F_{A-1,A}^{f,D}({\vec P}_{A-1})= \langle{\vec
P}_{A-1},\Psi_{A-1}^{f}S_G|\Psi_A^0\rangle\nonumber\\
&=&\int e^{i{\vec p}_{A-1}{\vec r}_1}S_G^{\dagger}({\vec r}_1\dots{\vec
r}_A){\Psi_{A-1}^f}^*({\vec r}_2\dots
{\vec r}_A)\Psi_A^0({\vec r}_1, {\vec r}_2\dots{\vec r}_A)\ 
\delta\left({\sum_{j=1}^A
{\vec r}_j}\right)\,\prod_{i=1}^Ad{\vec r}_i\ .
\label{overlap}
 \eeqn
 The quantity $S_G$ in Eq.~(\ref{overlap}) is the Glauber operator, which
describes the FSI of the debris from the struck proton with the $(A-1)$
system, i. e.
 \beq
S_G({\vec r}_1\dots{\vec r}_A)=
{\prod_{i=2}^A}\left[1 - \Gamma^{N^*N}(\vec b_1-\vec b_i,z_i-z_1)
\Theta(z_i-z_1)\right]\ ,
\label{SG}
 \eeq
 where ${\vec b}_i$ and $z_i$ are the transverse and longitudinal
components, respectively, of the coordinate of nucleon ``i", ${\vec
r}_i\equiv({\vec b}_i,z_i)$, ${\mit\Gamma}^{N^*N}({\vec b})$ is the
Glauber profile function describing the elastic scattering of the debris,
denoted $N^*$, with the nucleons of the $(A-1)$ system, and the function
$\theta(z_i-z_1)$ takes care of the fact that debris of the struck proton
``1'' propagates along a straight-path trajectory, so that they interacts
with nucleon ``$j$'' only if $z_j>z_1$. We have chosen the longitudinal
axis $z$ along the momentum of the virtual photon, and besides the usual
dependence on the transverse separation $\vec b_1-\vec b_i$ between $N^*$
and $N_i$ for a high-energy amplitude, we have also reserved a dependence
of $\Gamma^{N^*N}$ on the longitudinal separation $z_i-z_1$, which should
take care of the time dependence of the effective cross section discussed
above. Let us eventually point out that because of the effects from the
FSI, the Distorted Spectral Function depends now upon the vector ${\vec
P}_{A-1} \neq {\vec p}_1$. With the FSI taken into account, the Distorted
cross section is now Eq. \ref{crosa-1} with $P^A(|{\vec P_{A-1}}|,E)$
replaced by $P^{A,D}({\vec P_{A-1}},E)$. The cross section integrated
over the removal energy becomes proportional to the {\it distorted
momentum distribution}
 \beq 
n_D^A({\vec P}_{A-1})=\sum_f\left | F_{A,A-1}^{f,D}({\vec
P_{A-1}})\right |^2 ={(2 \pi)^{-3}} \int e^{i {\vec P}_{A-1}({\vec r}
-{\vec r}\,')}\rho_D ({\vec r},{\vec r}\,') d{\vec r} d{\vec r}'\ ,
  \label{nd1}
 \eeq
 where
 \beq
\rho_D ({\vec r},{\vec r}\,')=\int {{\Psi_A^0}^{*}({\vec r},{\vec
r}_2\dots{\vec r}_A)
S_G^{\dagger}({\vec r},{\vec r}_2\dots{\vec r}_A) S_G({\vec r}\,',{\vec
r}_2\dots{\vec r}_A)
\Psi_A^0({\vec r}\,',
{\vec r}_2\dots{\vec r}_A)}\prod_{i=2}^Ad{\vec r}_i
\label{rho_D}
 \eeq
 is the  one-body {\it distorted} mixed density matrix.

A general approach to calculate $\rho_D ({\vec r},{\vec r}\,')$ has been
developed in Ref. \cite{cd} in terms of correlated wave functions. Since
we are interested in the low momentum part of the wave function, and also
due to the exploratory nature of our calculations, we limit ourself here
to a traditional Glauber-type nuclear structure approach in which
$\Psi_A^0 ({\vec r},{\vec r}_2\dots{\vec r}_A)$ is written as a product
of a function $\phi$, describing the motion of nucleon ``1" and the wave
function $\Psi_{A-1}^f({\vec r}_2\dots{\vec r}_A)$ of the spectator, and
$|\Psi_{A-1}^f({\vec r}_2\dots{\vec r}_A)|^2$ is factorized into a
product of single particle densities. One obtains, in this case,
 \beq
n_D^A(\vec P_{A-1})\equiv N(\vec P_{A-1})=\left| F_{A,A-1}^{D}({\vec
P_{A-1}})\right |^2\ ,
\label{nondi}
 \eeq
 with
 \beqn
F_{A,A-1}^{D}({\vec P_{A-1}})
\simeq \int
e^{i{\vec P}_{A-1}\,\vec r}\,
\phi ({\vec r})
\left[1 - \frac{S({\vec b},z)}{2(A-1)}\right]^{A-1}\, 
d {\vec r}\ .
\label{b.8b}
 \eeqn
and
 \beq
S({\vec b},z)=\int\limits_{z}^{\infty} dz'\,
\rho_A({\vec b},z')\,\sigma_{eff}(z'-z)\ ,
\label{b.9}
 \eeq
 \noindent where $\rho_A({\vec b},z)$ is the nuclear density (normalized as $\int d^3r\,
\rho_A(\vec r)=A$), $\phi(\vec r)$ describes the relative motion of the
struck nucleon with respect to the spectator $A-1$ nucleus, and
$\sigma_{eff}(z'-z)$ is given by (\ref{b.5}) or (\ref{bb.6}).

We have estimated the Distorted form factor, Eq.~(\ref{b.9}),
within  the optical approximation, obtaining
 \beqn
F_{A,A-1}^{D}(\vec P_{A-1})&=&\int d^2b\,
e^{i\,{\vec P}_{\perp}\,\vec b}\,
\int\limits_{-\infty}^{\infty} dz\,
e^{i\,P_{l}z}\,
\phi(\vec b,z)\,
{\rm exp}\left[- \frac{1}{2}\,S(\vec b,z)\right]\ ,
\label{b.13}
 \eeqn
 where ${\vec P}_T$ and $P_L$ denote the transverse and longitudinal 
components of $\vec P_{A-1}$ with respect to $\vec q$. The optical
approximation is a rather good one for heavy nuclei, but it should be
abandoned in case of light nuclei for which Eq.~(\ref{rho_D}) has  to be
used.

The total cross section of the process is proportional to the integral of
the distorted transition matrix element, obtaining
 \beq
\sigma_{tot} \propto \int d^3P_{A-1}\,
\left |{F^D_{A,A-1}(P_{A-1})}\right |^2 =
\int d^2b\,\int\limits_{-\infty}^{\infty} dz\,
\rho (\vec b,z)\,
{\rm exp}\left[ -\,S(\vec b,z)\right]\ .
\label{b.10}
 \eeq

\section{Results of the calculations of the total cross section
and the distorted momentum distributions}

\begin{figure}[tbh] 
\includegraphics{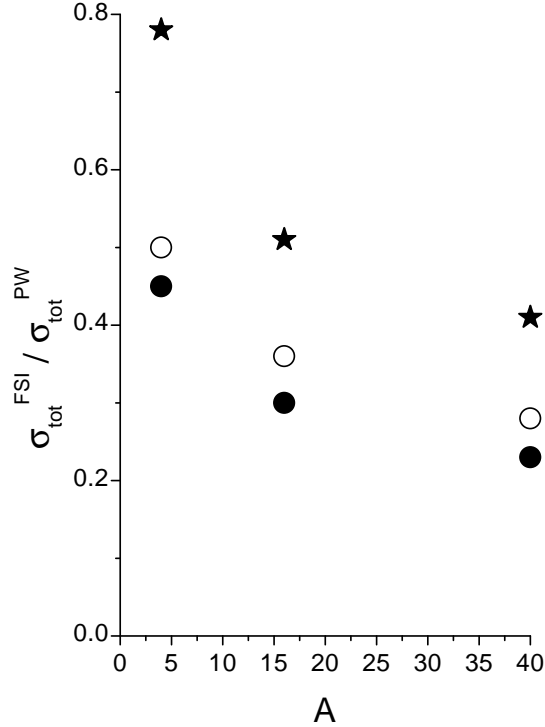}
\begin{center}                  
\vspace{10cm}
\parbox{13cm}
{\caption[Delta]
 {The ratio between the total cross section (Eq.~(\ref{b.10}) calculated taking
 into account ($S(\vec b, z) \neq 0$, $\sigma_{tot}\equiv \sigma_{tot}^{FSI}$)
 and disregarding 
 ($S(\vec b, z) = 0$, $\sigma_{tot}\equiv \sigma_{tot}^{PW}$) the Final State Interaction. 
  The open
dots correspond to the effective cross section of the nucleon debris given by
the color string model [Eq.~(\ref{b.5})], whereas the full dots
correspond to the cross section where the gluon bremsstrahlung has also
been considered [Eq.~(\ref{bb.6})] at $Q^2=100GeV^2$. The stars represent
the {\it proton} transparency calculated in Refs.~\cite{cd,hiko} for the
reaction $A(e,e'p)X$. }
 \label{Fig4}}
\end{center}
 \end{figure}

\begin{figure}[tbh] 
\includegraphics{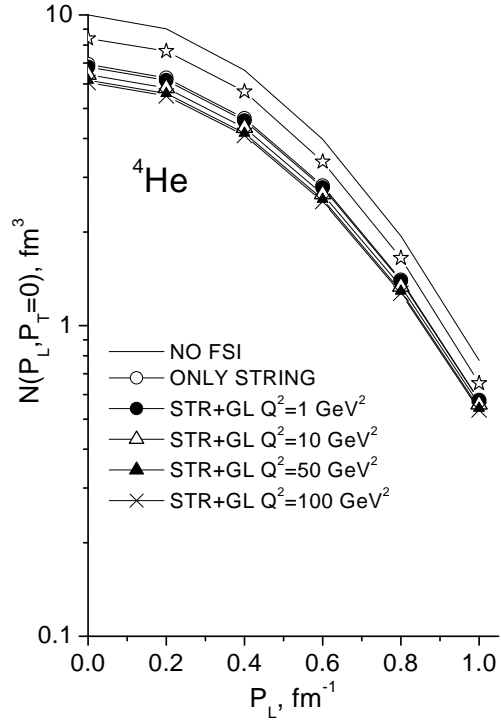}
\begin{center}                  
\vspace{10cm}
\parbox{13cm}
{\caption[Delta]
 {The proton Momentum Distribution for $^4He$ shown in Fig.1 (NO FSI)
compared with the Distorted Momentum Distribution $N({\vec P}_{A-1}$)  
(Eq.(\ref{nondi})) plotted {\it vs} $P_L$ for $P_T=0$. The
curve labeled by open dots has been obtained using the effective
 cross section for the nucleon debris corresponding to the color string model
(Eq. (\ref{b.5})), whereas the other curves correspond to the cross
section  which includes also the gluon
bremsstrahlung (Eq.(\ref{bb.6})). The stars represent the distorted {\it proton} momentum
distributions calculated in Ref. \cite{hiko} for the semi-inclusive process
$^4He(e,e'p)X$ .}
\label{Fig5}}
\end{center}
 \end{figure}

\begin{figure}[tbh] 
\includegraphics{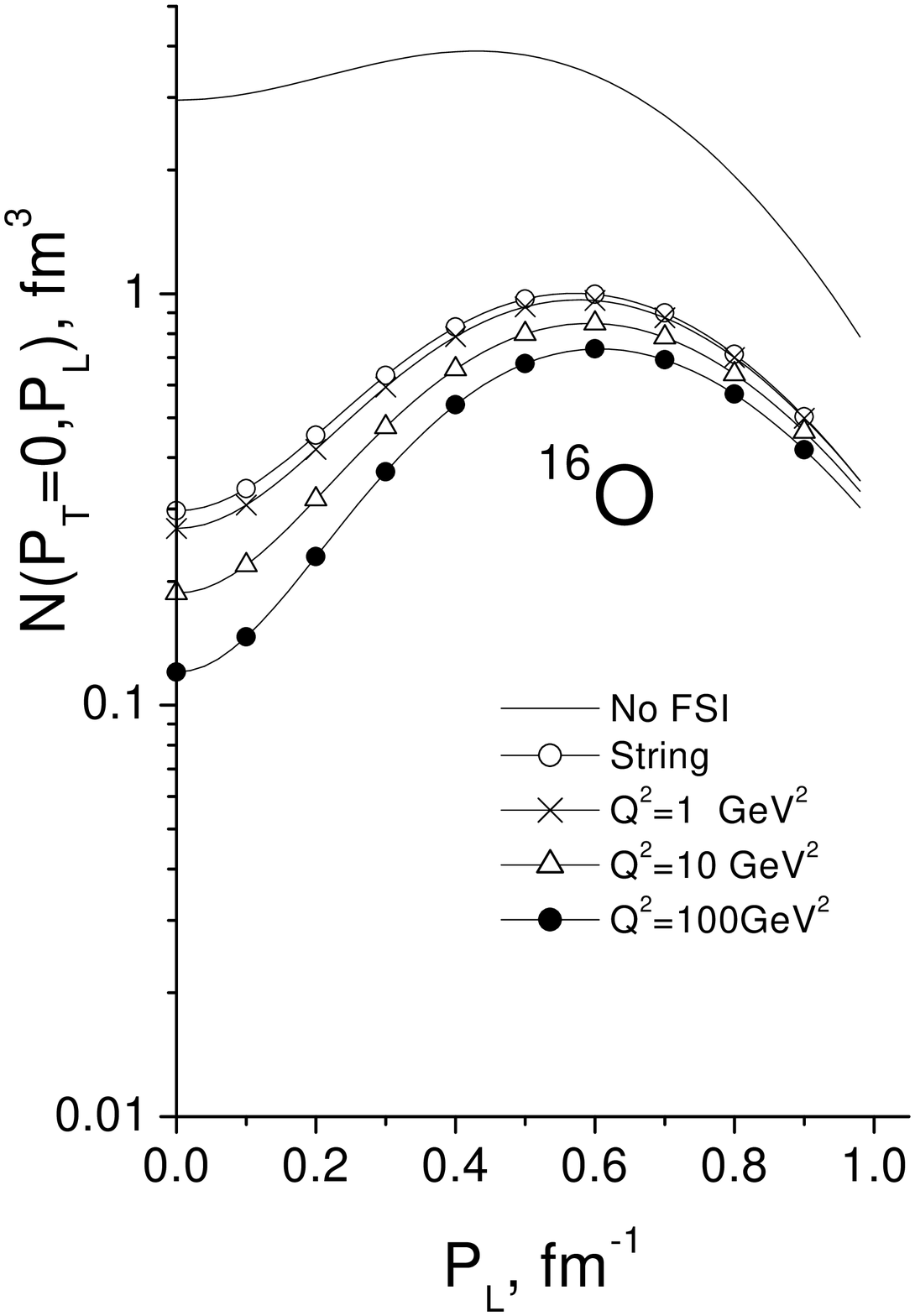}
\begin{center}                  
\vspace{10cm}
\parbox{13cm}
{\caption[Delta]
 {The same as in Fig.\ref{Fig5} but for $^{16}O$. The {\it proton}
rescattering in the reaction $^{16}O(e,e'p)X$ damps the momentum
distribution (full curve) by a factor of about 0.5 \cite{hiko}.}
 \label{Fig6}}
\end{center}
\end{figure}

\begin{figure}[tbh] 
\includegraphics{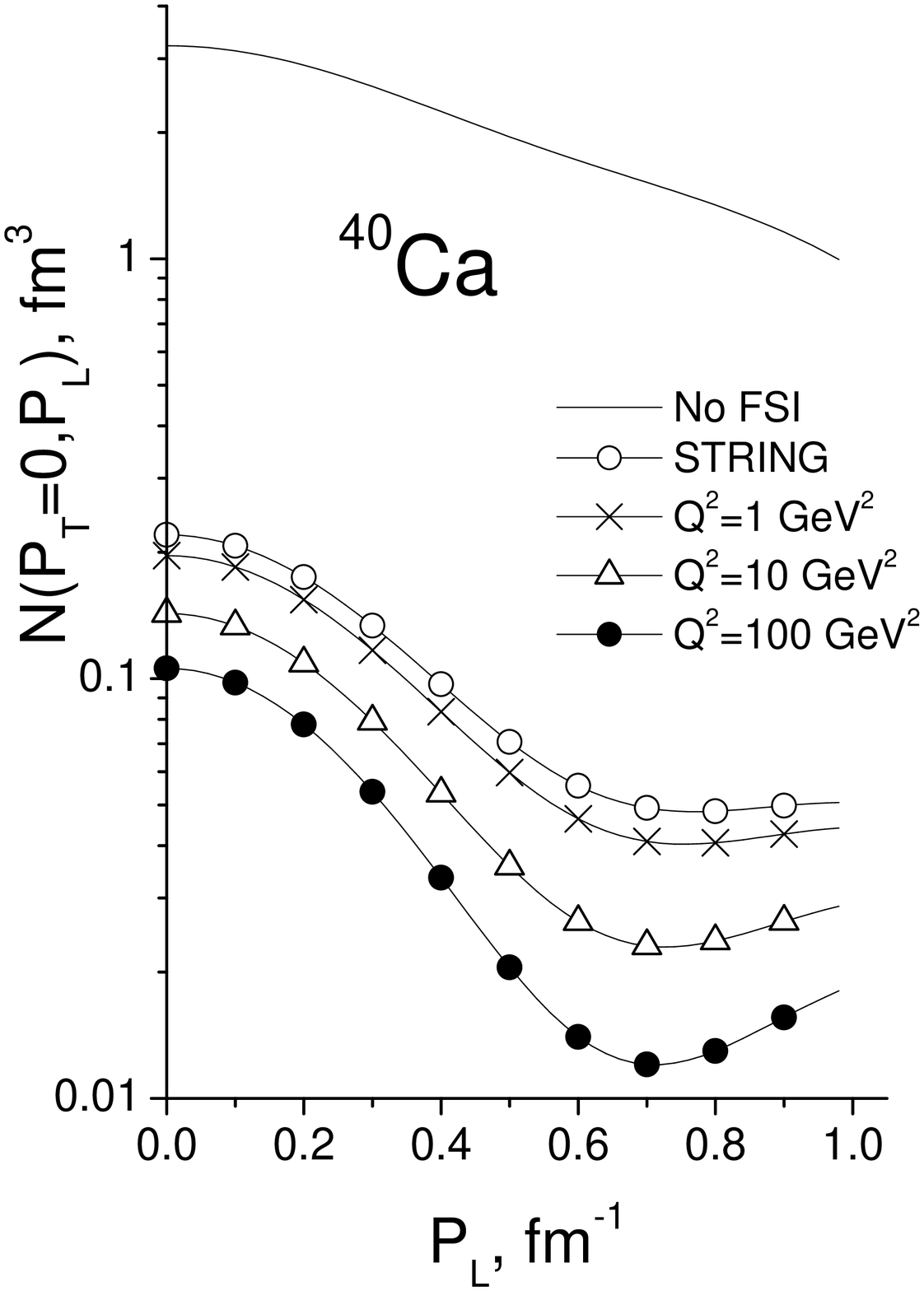}
\begin{center}                  
\vspace{10cm}
\parbox{13cm}
{\caption[Delta]
 {The same as in Fig. \ref{Fig4} but for $^{40}Ca$. The {\it proton}
rescattering in the reaction $^{40}Ca(e,e'p)X$ damps the momentum
distribution (full curve) by a factor of about 0.5 \cite{hiko}. .}
 \label{Fig7}}
\end{center}
 \end{figure}

\begin{figure}[tbh] 
\includegraphics{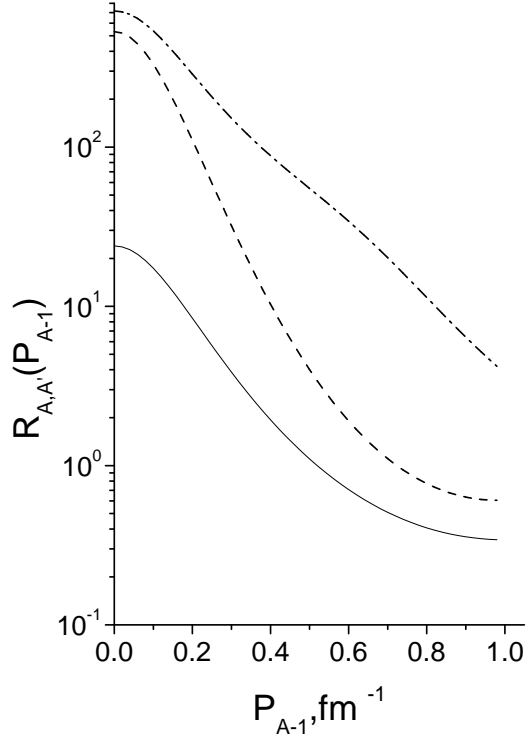}
\begin{center}                  
\vspace{10cm}
\parbox{13cm}
{\caption[Delta]
 {The same quantity as in Fig. \ref{Fig2} obtained with the Distorted
Momentum Distributions shown in Figs.\ref{Fig4}-\ref{Fig6} (STRING).
   Dotted line: $^4He$; dashed line: $^{16}O$; dot-dashed line:
$^{40}Ca$.}
 \label{Fig8}}
\end{center}
\end{figure}

\begin{figure}[tbh] 
\includegraphics{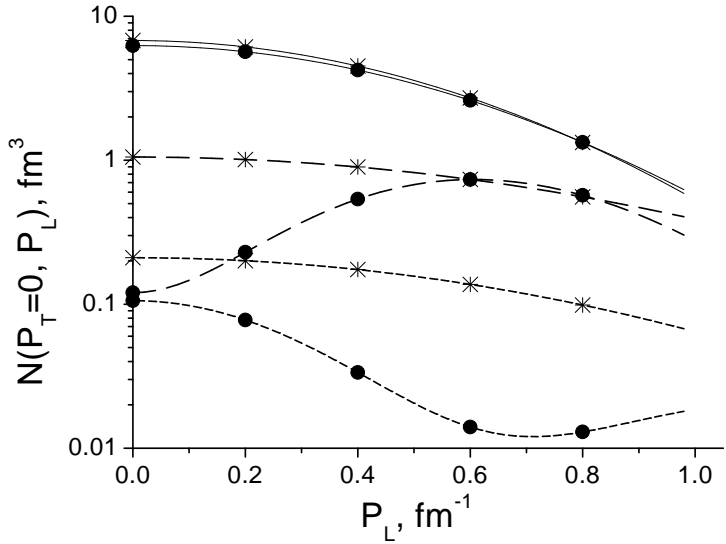}
\begin{center}                  
\vspace{7.5cm}
\parbox{13cm}
 {\caption[Delta] 
{The distorted momentum distributions for $^4He$
(full), $^{16}O$ (long dashes) and $^{40}Ca$ (short dashes), calculated
with two different forms for the function $\phi (\vec b,z)$ which
describe in the target nucleus the relative motion between the hit
nucleon and the spectator nucleus $(A-1)$ (cf. Eq. (\ref{b.8b}).  
Gaussian form: open dots; realistic form used in this paper: stars.}
 \label{Fig9}}
\end{center}
\end{figure}

We have calculated both the total cross section and the distorted
momentum distribution of the process $A(e,e'(A-1))X$ off \, $^4He$, $^{16}O$, and $^{40}Ca$;
the hit nucleon is considered to be a proton. For the
nuclear density $\rho_A(\vec b,z)$ we have used both the Harmonic
Oscillator and the Fermi distribution forms. As for the function $\phi
(\vec b_1, z_1)$,  which describes the relative motion of nucleon $1$ and
the spectator nucleus $A-1$, it has been chosen such that when $FSI$ are
absent, $N(|\vec P_{A-1}|)$ coincides with the low momentum part of the distributions
given in Ref. \cite{silva}. The parameters used for the calculation of
the effective cross sections [Eqs.~(\ref{b.5}) and (\ref{bb.6})] were as
follows: $\sigma_{tot}^{NN}=40\,\, mb$, $\sigma_{tot}^{\pi N}=20\,\, mb$,
$\Lambda_{QCD}=0.25\,\, GeV$, $\lambda = 0.65\,\, GeV$. The results of the
calculations are presented in Figs.~\ref{Fig4}--\ref{Fig9}.

Fig.~\ref{Fig4} shows the ratio of the distorted to undistorted total
cross sections defined by Eq.~(\ref{b.10}) calculated either taking into
account ($S(\vec b, z) \neq 0$), or disregarding ($S(\vec b, z) = 0$) the
effect of the nucleon debris  rescattering in the medium, respectively.\\

 As anticipated, the rescattering effects exhibit a decreasing
A-dependence. The effect of gluon radiation, responsible for the
difference between closed and open dots, is a nearly A-independent 10\%
correction. We have also calculated the nuclear transparency for the
reaction of quasi-elastic scattering $A(e,e'p)B$ where the intact struck
proton (Glauber approximation) propagates through the nucleus. These
results depicted by star points clearly exhibits the expected weaker
absorption effects for the proton compared to the debris in the process
$A(e,e'(A-1))X$.
The momentum distributions, Eq.~(\ref{n0}),  for $^4He$, $^{16}O$ and
$^{40}Ca$, are compared in Figs.~\ref{Fig5}--\ref{Fig7} with the distorted
momentum distributions defined by  Eq.~(\ref{nondi}) calculated for $P_T=0$.\\

 Concerning the results presented in these Figures, the following remarks
are in order:
 \begin{enumerate}
 \item 
 Due to the smaller dimensions, rescattering effects are less important
in $^4He$ than in heavier nuclei. As a matter of fact, in the former
case, they simply reduce, in the considered range of momenta, to an
almost constant reduction of about 50 $\%$, whereas it appears that for
medium weight and heavy nuclei not only the absorption is substantially
stronger, but appreciable distortion effects are apparent, with the role
of gluon radiation increasing with $A$.
 \item 
 It is interesting to compare the process $A(e,e'(A-1))X$ we are
investigating occurring at Bjorken $x \ll 1$, with the semi-inclusive {\it
nucleon} knock-out $A(e,e'p)X$ occurring at Bjorken $x \simeq 1$ (note
that $X$ refers to the proton debris in the $A(e,e'(A-1))X$ process, and 
to a $A-1$ nucleon state in the $A(e,e'p)X$ process). Within the $PW$
approximation the two processes are proportional to the same nuclear
part, {\it viz} the nucleon momentum distribution $n^A(p)$ (Eq.
(\ref{nk})), whereas when the FSI is considered the first process will be
distorted by the nucleon {\it nucleon debris} rescattering, and the second one by
the {\it proton} rescattering. Thus, it appears that by comparing the two
processes the differences between nucleon debris and proton propagations
in the nuclear medium can be investigated. The calculation of
rescattering effects in the processes $^4He(e,e'p)X$, $^{16}O(e,e'p)X$
and $^{40}Ca(e,e'p)X$ has been performed in \cite{hiko} by a Glauber-type
approach and the results can be summarized as follows:  in the momentum region considered in
this paper,
 the effects of the FSI, due to the proton rescattering,  damp the momentum distributions by an almost
momentum-independent amount, the reduction factors being 0.85, 0.5 and
0.5 in $^4He$, $^{16}O$ and $^{40}Ca$, respectively. By comparing these
results with the ones presented in Figs.~\ref{Fig4}--\ref{Fig6}, we can
conclude that, apart from $^4He$, the effects of proton and nucleon
debris propagation are, as expected, very different.
 \end{enumerate}

In Fig. \ref{Fig8} the ratio $R_{A,A'}$ (Eq. (\ref{eq5})), which is
plotted in Fig. \ref{Fig2} in the case of the $PW$ approximation, is
shown when the nucleon debris rescattering  is taken into account in the final state. 
 It is gratifying to see that when the final state rescattering of the
debris is taken into account, the differences between different nuclei,
i.e. nuclear effects, are even emphasized.

In Fig. \ref{Fig9}, the dependence of our results upon the choice of the
function $\phi(\vec b,z)$ describing in the target ground state the
relative motion between the hit nucleon and the spectator $A-1$ , is
exhibited by comparing the results obtained with the realistic function
used in the calculations with the results obtained with a simple Gaussian
function. It can be seen that, apart from $^4He$, the use of a Gaussian
function is not recommended.

\begin{figure}[tbh] 
\includegraphics{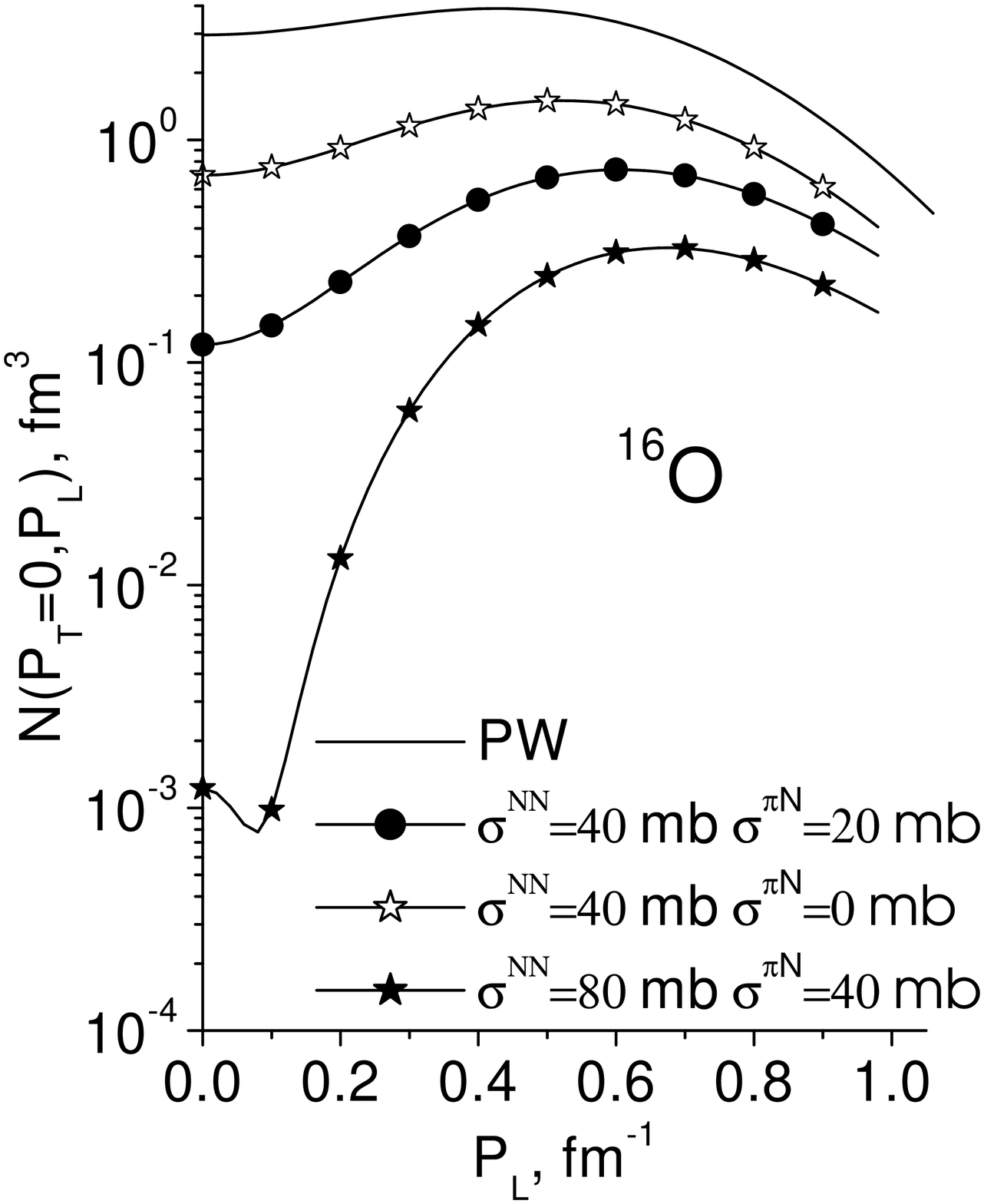}
\begin{center}                  
\vspace{10cm}
\parbox{13cm}
 {\caption[Delta] {The sensitivity of the distorted momentum
distributions of $^{16}O$ upon the values of the Nucleon-Nucleon and
Pion-Nucleon elementary cross sections. The full curve represents the
Plane Wave approximation,whereas  the other curves were obtained using the debris
effective cross section given by Eq.\ref{bb.6} .}
 \label{Fig10}}
\end{center}
\end{figure}

Eventually, in Fig. \ref{Fig10}, the sensitivity of our results upon the
quantities entering the effective cross section (Eq. \ref{bb.6}), {\it
viz} $\sigma^{NN}_{tot}$, $\sigma^{\pi N}_{tot}$ and $\lambda$, is
illustrated.

\section{Summary and Conclusions}

In this paper we have considered a theoretical model to describe the FSI
arising from the propagation and hadronization of the partonic debris
produced from the hard scattering of a lepton off a nucleon bound in a
nucleus $A$. The FSI arises due to the rescattering of the hadrons which
are formed both from string breaking and from gluon radiations. In order
to experimentally investigate the correctness of the model, we have
analyzed the semi-inclusive process in which, instead of a secondary,
e.g. leading hadron, the whole $A-1$ nucleus in low momentum and energy
states, is detected in coincidence with the scattered lepton. In absence
of any FSI, the momentum distributions of the $A-1$ nucleus would be
nothing but the momentum distributions of the hit nucleon and its
integral will give the total number of surviving $A-1$ nuclei; when the
FSI is switched on, the surviving probability will be reduced and the
momentum distributions will be distorted. We have shown that both effects
sensibly depend upon the details of the rescattering and hadronization of
the nucleon debris. By comparing the results for the process
$A(e,e'(A-1))X$ with the much investigated process of nucleon knock-out,
$A(e,e'p)X$, the differences between the {\it nucleon} and its {\it
debris} propagations in the nuclear environment can be investigated. For
example, we have found that whereas 60\% of the knocked out protons off
$^{16}O$ at $Q^2 = GeV^2$ escape the nucleus without interaction, for a
debris of a struck nucleon the survival probability reduces down to 25\%.
Furthermore, the momentum distributions of the recoiling $A-1$ nuclei
turn out to be very sensitive to the details of the model for
hadronization in nuclear environment. Up to now, information on hadron
formation time and hadronization, has been collected mainly from the
measurement of the ratio between the semi-inclusive cross section for
leading hadron production for a nucleus $A$ to that for the deuteron.
This ratio increases with energy approaching 1 when the formation time of
leading hadrons exceeds the nuclear size. In the process we are
proposing, the effects of FSI do not vanish with energy, and vary from a
factor of 2-3 in a light nucleus like $^4He$, to orders of magnitudes in
heavy nuclei. Moreover, they generate peculiar and strong distortions of
the nucleon momentum distributions. Thus, from a theoretical point of
view the semi-inclusive process $A(e,e'(A-1))X$ appears to be very
promising, for nuclear effects not only manifest themselves in a relevant
quantitative way, but also produce peculiar qualitative effects.

An essential advantage of the process under discussion, compared to
inclusive hadron production, is the possibility to study the early stage of
hadronization at short formation times without being affected by
cascading processes. Indeed, no cascading is possible if the recoil
nucleus $(A-1)$ survives. At the same time,  most of hadrons with small
momentum produced in inclusive process $A(e,e'h)X$ originate from
cascading of more energetic particles. In order to analyze data and
extract information on the early stage of hadronization, one needs a
realistic model for cascading what is barely possible.
 
 \section{Acknowledgments} We are grateful to D. Treleani for several
fruitful discussions during the first stage of this work and to L.
Kaptari for useful suggestions. CdA is indebted to B. Povh and the
Max-Planck Institute f\"ur Kernphysik, Heidelberg, for several
invitations which made the completion of this work possible.  This work
has partly been performed under the contract HPRN-CT-2000-00130 of the
European Union and supported by the grant from the Gesellschaft f\"ur
Schwerionenforschung Darmstadt (GSI), grant No.~GSI-OR-SCH, and by the
European Network {\it Hadronic Physics with Electromagnetic Probes},
Contract No.~FMRX-CT96-0008. Partial support by the Ministero
dell'Istruzione, Universit\`a e Ricerca (MIUR), through the funds
COFIN01, is also acknowledged.

\end{document}